\documentclass[preprint,12pt]{elsarticle}

\usepackage{bm}
\usepackage{amsmath,amssymb,amsfonts,times}
\usepackage{url}

\usepackage[colorlinks=true, pdfstartview=FitV, linkcolor=red, citecolor=blue, urlcolor=blue]{hyperref}

\usepackage{graphicx}
\usepackage{epsfig}
\usepackage{slashed}
\usepackage{a4wide}
\usepackage{fancyhdr}
\usepackage{subfigure}

\usepackage{rotating}

\usepackage{pdflscape}

\usepackage{multirow}

\allowdisplaybreaks[1]
\newcommand{\UNIT}[1]{\mbox{$\,{\rm #1}$}}
\newcommand{\MeV}{\UNIT{MeV}}
\newcommand{\GeV}{\UNIT{GeV}}

\newcommand{\ds}{\displaystyle}
\newcommand{\zerovec}{\vec{0}\,}

\newcommand{\pvec}{\vec{p}\,}

\newcommand{\Gcapvec}{\vec{G}\,}

\newcommand{\sigmavec}{\vec{\sigma}\,}

\newcommand{\tauvec}{\vec{\tau}\,}
\newcommand{\rhovec}{\vec{\rho}\,}
\newcommand{\deltavec}{\vec{\delta}\,}

\newcommand{\be}{\begin{equation}}
\newcommand{\ee}{\end{equation}}
\newcommand{\ba}{\begin{eqnarray}}
\newcommand{\ea}{\end{eqnarray}}

\newcommand{\fac}{\frac{\kappa}{(2\pi)^{3}}}

\newcommand{\nld}{{\cal D}}

\newcommand{\nldl}{\overleftarrow{{\cal D}}}
\newcommand{\nldr}{\overrightarrow{{\cal D}}}

\newcommand{\pspace}{\int\limits_{|\pvec|\leq p_{F_{B}}}\!\!\!\!\!\! d^{3}p}

\newcommand{\partialr}{\overrightarrow{\partial}}
\newcommand{\partiall}{\overleftarrow{\partial}}

\newcommand{\xil}{\overleftarrow{\xi}}
\newcommand{\xir}{\overrightarrow{\xi}}

\newcommand{\Bla}{\Big<}
\newcommand{\Bra}{\Big>}
\begin{document}

\begin{frontmatter}

\title{Momentum dependent mean-fields of (anti)hyperons}
%\title{Momentum dependent $\Lambda$ ($\overline{\Lambda}$)-, 
%$\Sigma$ ($\overline{\Sigma}$)- and $\Xi$ ($\overline{\Xi}$)-dynamics}

\author{T.~Gaitanos, A. Chorozidou}
\address{Department of Theoretical Physics, 
Aristotle University of Thessaloniki, GR-54124 Thessaloniki, Greece}
\address{email: tgaitano@auth.gr}

\begin{abstract}
We investigate the in-medium properties of hyperons and anti-hyperons in the 
framework of the Non-Linear Derivative (NLD) model. We focus on the momentum 
dependence of in-medium strangeness optical potentials. 
The NLD model is based on the simplicity of the well-established 
Relativistic Mean-Field (RMF) approximation, but it incorporates an explicit 
momentum dependence on a field-theoretical level. 
The extension of the NLD model to the (anti)baryon-octet is 
formulated in the spirit of SU(6) and G-parity arguments. 
%It is shown that the 
%momentum cut-offs force the $\Lambda$, $\Sigma$ and $\Xi$ optical potentials 
%to be consistent with recent studies of the chiral effective field theory and 
%Lattice-QCD calculations over a wide momentum region. 
%%%%%%%%%%%%%%%%%%%%%%%%%%% changes in revision - minor 1
It is shown that with an appropriate choice of momentum cut-offs the 
$\Lambda$, $\Sigma$ and $\Xi$ optical potentials are 
consistent with recent studies of the chiral effective field theory and 
Lattice-QCD calculations over a wide momentum region. 
%%%%%%%%%%%%%%%%%%%%%%%%%%% changes in revision - minor 1 
In addition, we present NLD predictions for the in-medium momentum dependence of 
$\overline{\Lambda}$-, $\overline{\Sigma}$- and $\overline{\Xi}$-hyperons. 
This work is important for future experimental studies such as CBM, PANDA at 
the Facility for Antiproton and Ion Research (FAIR). It is relevant for 
nuclear astrophysics too. 
\end{abstract}

\begin{keyword}
Equations of state of hadronic matter, optical potential, in-medium hyperon potentials.
\end{keyword}

\end{frontmatter}

\date{\today}

%%%%%%%%%%%%%%%%%%%%%%%%%%%%%%%%%%%%%%%%%%%%%%%%%%%%%%%%%%%%%%%%%%%%%%%%%%%%%%%
\section{\label{sec1}Introduction}
%%%%%%%%%%%%%%%%%%%%%%%%%%%%%%%%%%%%%%%%%%%%%%%%%%%%%%%%%%%%%%%%%%%%%%%%%%%%%%%

Astrophysical observations on particularly massive neutron 
stars~\cite{NS1,NS2,NSnew} 
have driven the nuclear physics and astrophysics communities 
to detailed investigations 
of the nuclear equation of state (EoS) under conditions far beyond 
the ordinary matter~\cite{Lattimer14}. On one hand, 
theoretical and experimental studies on heavy-ion 
collisions over the last few decades concluded a softening of the high-density EoS 
in agreement with phenomenological and 
microscopic models~~\cite{softeos1,softeos4,softeos5}. On the other hand, the 
observations of two-solar mass pulsars~\cite{NS1,NS2,NSnew} together with additional 
constraints on the high-density limit of the speed of sound~\cite{babis} 
gave some controversial insights on the EoS of compressed baryonic matter. 
They provide an upper limit for the neutron star mass by excluding 
soft-type hadronic EoS's at high baryon densities. 

Compressed baryonic matter may consist not only of nucleons. It can include 
fractions of heavier baryons, when their production is energetically allowed. 
These are the hyperons $\Lambda,~\Sigma$  and $\Xi$ 
%%%%%%%%%%%%%%%%%%%%%%%%%%% changes in revision - minor 2
%and $\Omega$ 
%%%%%%%%%%%%%%%%%%%%%%%%%%% changes in revision - minor 2
as  a part of 
the irreducible representations of SU(3). While the nucleon-nucleon (NN) interaction 
is very well known, the hyperon interactions are still not fully 
understood. Indeed, there are many experimental data for NN-scattering in free space and 
inside hadronic media (finite nuclei, heavy-ion collisions, hadron-induced reactions) 
allowing a precise determination of the NN-interaction. 
%However, due to the 
%scarce experimental access to the strangeness sector (hyperon-nucleon (YN) or 
%hyperon-hyperon (YY) interactions) the YN-potentials are still theoretically little 
%understood. 
%As a consequence, theoretical predictions to the in-medium Y-interactions are considered 
%so far as an extrapolation 
%to high densities. 
%%%%%%%%%%%%%%%%%%%%%%%%%%% changes in revision - minor 3
Concerning the strangeness sector (hyperon-nucleon (YN) or hyperon-hyperon (YY) 
interactions), there exist phenomenological and microscopic models with predictions 
for the in-medium hyperon properties at matter densities close to saturation and at higher 
densities. However, the experimental access to the strangeness sector is still scarce. 
%%%%%%%%%%%%%%%%%%%%%%%%%%% changes in revision - minor 3
A common prediction of theoretical models is a considerable softening 
of the hadronic EoS at high densities by adding to a system more degrees of freedom such as 
strangeness particles. The inclusion of hyperons into nuclear approaches made many 
of them, which were successfully applied to nuclear systems (nuclear matter, finite nuclei, 
nuclear reactions), incompatible with the astrophysical observations of two-solar mass 
pulsars~\cite{NS1,NS2}. This is the so-called hyperon-puzzle~\cite{puzzle,puzzle1}. This 
puzzle has received recently theoretical attraction by a new observation of a quite massive 
neutron star~\cite{NSnew}. A comprehensive theoretical view concerning the 
microscopic descriptions of in-medium properties of the baryon-octet is given in 
Ref.~\cite{review}. There exist also theoretical reviews based on the RMF approximation, 
see for instance Refs.~\cite{rmf-others-1,rmf-others-2,rmf-others-3}.

It is thus of great interest to address the in-medium behaviour of hyperons in nuclear 
matter, as we do in this work. We use an alternative RMF approach based on the fact, 
that compressed matter consists of particles with high relative momenta. Therefore, 
not only the density dependence, but 
the momentum dependence of the in-medium interactions is important too. 
The reason for doing 
so is that conventional RMF-models do not explain the empirical saturation of the 
in-medium interactions of high-momenta (anti)nucleons. In terms 
of SU(6) this issue appears for high-momenta (anti)hyperons too. This is the 
Non-Linear Derivative (NLD) model~\cite{nld}. It retains the basic RMF Lagrangian 
formulation, but it includes higher-order derivatives in the NN-interaction Lagrangians. 
It has been demonstrated that this Ansatz corrects 
the high-momentum behaviour of the interaction, makes the EoS softer at densities just 
above saturation, but at the same time it reproduces the two-solar mass pulsars at 
densities far beyond saturation~\cite{nld}. Here we extend the NLD approach by 
including strangeness into the nuclear matter and discuss the momentum dependence 
of the in-medium hyperon potentials. 

%%%%%%%%%%%%%%%%%%%%%%%%%%%%%%%%%%%%%%%%%%%%%%%%%%%%%%%%%%%%%%%%%%%%%%%%%%%%%%%%%%%
\section{\label{sec2}The NLD Model for the baryon octet}
%%%%%%%%%%%%%%%%%%%%%%%%%%%%%%%%%%%%%%%%%%%%%%%%%%%%%%%%%%%%%%%%%%%%%%%%%%%%%%%%%%%

In this section we briefly introduce the non-linear derivative (NLD) model and 
extend it to the baryon octet. A detailed description of the NLD model for nucleons 
can be found in Ref.~\cite{nld}. The NLD-Lagrangian is based on the 
conventional Relativistic Hadro-Dynamics (RHD)~\cite{rhd} and it reads as 
%%%%%%%%%%%%%%%%%
\begin{align}
{\cal L} = & \frac{1}{2}
\sum_{B}
\left[
	\overline{\Psi}_{B}\gamma_{\mu} i\partialr^{\mu}\Psi_{B}
	- 
	\overline{\Psi}_{B} i\partiall^{\mu} \gamma_{\mu} \Psi_{B}
\right]
- \sum_{B}m_{B} \overline{\Psi}_{B}\Psi_{B}
\nonumber\\
- &
\frac{1}{2}m^{2}_{\sigma}\sigma^{2}
+\frac{1}{2}\partial_{\mu}\sigma\partial^{\mu}\sigma
-U(\sigma)
\nonumber\\
+ & \frac{1}{2}m^{2}_{\omega}\omega_{\mu} \omega^{\mu} 
-\frac{1}{4}F_{\mu\nu}F^{\mu\nu}
\nonumber\\
+ &
\frac{1}{2}m^{2}_{\rho}\rhovec_{\mu}\rhovec^{\mu} 
-\frac{1}{4}\Gcapvec_{\mu\nu}\Gcapvec^{\mu\nu}
-\frac{1}{2}m^{2}_{\delta}\deltavec^{2}
+\frac{1}{2}\partial_{\mu}\deltavec \, \partial^{\mu}\deltavec
\nonumber\\
+ & 
{\cal L}_{int}^{\sigma}+{\cal L}_{int}^{\omega}+
{\cal L}_{int}^{\rho}+{\cal L}_{int}^{\delta}
\label{NDC-free}
\,.
\end{align}
%%%%%%%%%%%%%%%%%
%\end{widetext}
The sum over $B$ runs over the baryonic octet
%%%%%%%%%%%%%%%%%
\begin{align}
\Psi_{B}= & (\Psi_{N}, \Psi_{\Lambda}, \Psi_{\Sigma}, \Psi_{\Xi} )^{T}
\label{Psitot}
\end{align}
%%%%%%%%%%%%%%%%%
with
%%%%%%%%%%%%%%%%%
\begin{align}
\Psi_{N}= & (\psi_{p},\psi_{n})^{T},~~
\Psi_{\Lambda}= \psi_{\Lambda}
\label{PsiL}\\
\Psi_{\Sigma}= & (\psi_{\Sigma^{+}},\psi_{\Sigma^{0}},\psi_{\Sigma^{-}})^{T},~~
\Psi_{\Xi}= (\psi_{\Xi^{0}},\psi_{\Xi^{-}})^{T}
\label{PsiX}
\end{align}
%%%%%%%%%%%%%%%%%
for the isospin-doublets $\Psi_{N}$ and $\Psi_{\Xi}$, isospin-triplet 
$\Psi_{\Sigma}$ and the neutral $\Psi_{\Lambda}$. 
The interactions between the nucleon fields are described by the exchange of
meson fields. These are the scalar $\sigma$ and vector $\omega^{\mu}$ mesons 
in the isoscalar channel, as well as the scalar $\deltavec$ and vector 
$\rhovec^{\mu}$ mesons in the isovector channel. Their corresponding 
Lagrangian densities are of the Klein-Gordon and Proca types, respectively. 
The term $U(\sigma)=\frac{1}{3}b\sigma^{3}+\frac{1}{4}c\sigma^{4}$ 
contains the usual selfinteractions of the $\sigma$ meson. 
The notations 
for the masses of fields in Eq.~(\ref{NDC-free}) are obvious. The field
strength tensors are defined as 
$F^{\mu\nu}=\partial^{\mu}\omega^{\nu}-\partial^{\nu}\omega^{\mu}$, 
$\Gcapvec^{\mu\nu}=\partial^{\mu}\rhovec^{\nu}-\partial^{\nu}\rhovec^{\mu}$ 
for the isoscalar and isovector fields, respectively. In the following 
we restrict to a minimal set of interaction degrees of freedom. In the 
iso-scalar sector, the $\sigma$- and $\omega$-fields are obviously considered. 
In the iso-vector channel, we keep the vector, iso-vector $\rho$-meson field and 
neglect the $\delta$-field. 

The NLD interaction Lagrangians contain the conventional RHD combinations 
between the bilinear baryon- and linear meson-fields, however, they are extended 
by the inclusion of non-linear derivative operators 
$\nldr, \nldl$ for each baryon species $B$:
%%%%%%%%%%%%%%%%%
%\begin{widetext}
\begin{equation}
{\cal L}_{int}^{\sigma} = \sum_{B} \frac{g_{\sigma B}}{2}
	\left[
	\overline{\Psi}_{B}
	\, \nldl_{B}
	\Psi_{B}\sigma
	+\sigma\overline{\Psi}_{B}
	\, \nldr_{B}
	\Psi_{B}
	\right]\,,
\end{equation}
\begin{equation}
{\cal L}_{int}^{\omega} = - \sum_{B} \frac{g_{\omega B}}{2}
	\left[
	\overline{\Psi}_{B}
	 \, \nldl_{B}
	\gamma^{\mu}\Psi_{B}\omega_{\mu}
	+\omega_{\mu}\overline{\Psi}_{B}\gamma^{\mu}
	\, \nldr_{B}
	\Psi_{B}
	\right]	\,,
\end{equation}
\begin{equation}
{\cal L}_{int}^{\rho} = - \sum_{B} \frac{g_{\omega \rho}}{2}
	\left[
	\overline{\Psi}_{B}
	 \, \nldl_{B}
	\gamma^{\mu}\vec{\tau}\Psi_{B}\rhovec_{\mu}
	+\rhovec_{\mu}\overline{\Psi}_{B}\vec{\tau}\gamma^{\mu}
	\, \nldr_{B}
	\Psi_{B}
	\right]	\,,
\label{NDCrd}
\end{equation}

%\end{widetext}
%%%%%%%%%%%%%%%%%
for the isoscalar-scalar, isoscalar-vector and isovector-vector vertices, 
respectively. The arrows on the non-linear operator $\nld_{B}$ indicate the direction 
of their action. The only difference with respect to 
the conventional RHD Lagrangian is the presence of additional
operator functions $\nldr_{B} ,~\nldl_{B}$. As we will see, they will regulate the 
high momentum component of hyperons. For this reason we will call them as 
regulators too. The operator functions (or regulators) $\nldr_{B},~\nldl_{B}$ are 
hermitian and generic functions of partial derivative operator. That is, 
$\nldr_{B} := \nld\left( \xir_{B} \right)$ and $\nldl_{B} := \nld \left( \xil_{B} \right)$ 
with the operator arguments $\xir_{B} = -\zeta_{B}^{\alpha}i\partialr_{\alpha},~
\xil_{B} = i\partiall_{\alpha}\zeta_{B}^{\alpha}$. The four vector 
$\zeta_{B}^{\mu}=v^{\mu}/\Lambda_{B}$ contains the cut-off $\Lambda_{B}$ and $v^{\mu}$ 
is an auxiliary vector. These regulators are assumed to 
act on the baryon spinors $\Psi_{B}$ and $\overline{\Psi}_{B}$ by a formal Taylor 
expansion with respect to the operator argument. The functional 
form of the regulators is constructed such that in the limit $\Lambda_{B}\to\infty$ 
the original RHD Lagrangians are recovered, that is, 
$\nldr_{B}=\nldl^{\dagger}_{B} \to 1$.

%%%%%%%%%%%%%%%%%%%%%%%%%%%%%%%%%%%%%%%%%%%%%%%%%%%%%%%%%%%%%%%%%%%%%%%%%%%%%%%%%%%
%\section{\label{sec4} NLD Field equations}
%%%%%%%%%%%%%%%%%%%%%%%%%%%%%%%%%%%%%%%%%%%%%%%%%%%%%%%%%%%%%%%%%%%%%%%%%%%%%%%%%%%

The presence of higher-order partial derivatives in the Lagrangian mediate a 
modification of the field-theoretical prescriptions. As discussed in detail in 
the original work of Ref.~\cite{nld}, the generalized Euler-Lagrange equations 
as well as the Noether-currents contain additional infinite terms of higher-order 
partial derivative contributions. However, the main advantage of the NLD approach 
relies on the fact that these terms can be resummed to compact expressions. 

From the generalized Euler-Lagrange formalism we obtain the equations of motion 
for the degrees of freedom in the NLD model. The meson field equations of motion 
read 
%%%%%%%%%%%%%%%%%
\begin{align}
&
\partial_{\alpha}\partial^{\alpha}\sigma + m_{\sigma}^{2}\sigma 
+ \frac{\partial U}{\partial\sigma} =
\frac{1}{2} \sum_{B} g_{\sigma B}
\left[
	\overline{\Psi}_{B} \, \nldl_{B} \Psi_{B} + 
	\overline{\Psi}_{B}\nldr_{B} \Psi_{B}
\right] \,,
\label{sigma_meson}\\
%\end{align}
%%%%%%%%%%%%%%%%%
%%%%%%%%%%%%%%%%%
%\begin{align}
&
\partial_{\mu}F^{\mu\nu} + m_{\omega}^{2}\omega^{\nu} =
\frac{1}{2} \sum_{B} g_{\omega B}
\left[
	\overline{\Psi}_{B}\, \nldl_{B} \gamma^{\nu}\Psi_{B} + 
	\overline{\Psi}_{B}\gamma^{\nu}\nldr_{B} \Psi_{B}
\right] \,,
\label{omega_meson}\\
%\end{align}
&
\partial_{\mu}G^{\mu\nu} + m_{\rho}^{2}\rhovec^{\nu} =
\frac{1}{2} \sum_{B} g_{\rho B}
\left[
	\overline{\Psi}_{B}\, \nldl_{B} \gamma^{\nu}\vec{\tau}\Psi_{B} + 
	\overline{\Psi}_{B}\vec{\tau}\gamma^{\nu}\nldr_{B} \Psi_{B}
\right] \,,
\label{rho_meson}
%%%%%%%%%%%%%%%%%
\end{align}
%%%%%%%%%%%%%%%%%
for the isoscalar-scalar, isoscalar-vector and isovector-vector exchange mesons, 
respectively. 

Each baryon-field obeys a Dirac-equation of the following type
%%%%%%%%%%%%%%%%%
\begin{equation}
\left[
	\gamma_{\mu}(i\partial^{\mu}-\Sigma^{\mu}_{B}) - 
	(m_{B}-\Sigma_{sB})
\right]\psi_{B} = 0
\;,
\label{Dirac_nld}
\end{equation}
%%%%%%%%%%%%%%%%%
with the selfenergies $\Sigma^{\mu}_{B}$ and $\Sigma_{sB}$ defined as 
%%%%%%%%%%%%%%%%%
\begin{eqnarray}
\Sigma^{\mu}_{B} & = & g_{\omega _{B}}\omega^{\mu}\nldr_{B} + 
g_{\rho B}\tauvec_{B} \cdot \rhovec^{\mu}\nldr_{B}~,
\label{Sigmav}\\
%\end{eqnarray}
%%%%%%%%%%%%%%%%%%
%%%%%%%%%%%%%%%%%%
%\begin{eqnarray}
\Sigma_{sB} & = & g_{\sigma B}\sigma\nldr_{B}
\;. \label{Sigmas}
\end{eqnarray}
%%%%%%%%%%%%%%%%%
Both Lorentz-components of the selfenergy, $\Sigma^{\mu}$ and $\Sigma_{s}$, 
show an explicit linear behaviour with respect to the meson fields $\sigma$, 
$\omega^{\mu}$ and $\rhovec^{\mu}$ as in the standard RHD. 
However, they contain an additional dependence on the regulators. General 
expressions for the Noether-current and energy-momentum tensor can also 
be derived. We give them below in the RMF approximation. 

The RMF application of the NLD formalism to static hadronic matter follows 
the same procedure as in the conventional RHD. The spatial components 
of the meson fields in Minkowski- and isospin-spaces vanish, 
$\omega^{\mu}\to (\omega^0,~\zerovec)$ and 
$\rhovec^{\mu} \to(\rho^0_3,~\zerovec_{3})$. For simplicity, we denote
in the following the remaining isospin component of the isovector fields as 
$\rho^{\mu}$. 
The solutions of the RMF equations start with the usual plane wave \textit{ansatz}
$\psi_{B}(s,\pvec) = u_{B}(s,\pvec)e^{-ip^{\mu}x_{\mu}} $ 
where $B$ stands for the various isospin states of the baryons and 
$p^{\mu}=(E,\vec{p}\,)$ is the single baryon 4-momentum. 
The application of the non-linear derivative operator $\nld_{B}$ to the 
plane wave \textit{Ansatz} of the spinor fields results in 
regulators $\nld_{B}$ which are now functions of 
the scalar argument $\xi_{B}=-\frac{v_{\alpha}p^{\alpha}}{\Lambda_{B}}$. That is, 
they depend explicitly on the single baryon momentum $p$ (with an appropriate 
choice of the auxiliary vector $v^{\alpha}$) and on the cut-off 
$\Lambda_{B}$, which may differ for each baryon type $B$. 
Each baryon fulfils a Dirac equation with the same form as in Eq.~(\ref{Dirac_nld}) and 
with corresponding explicitly momentum dependent scalar and vector selfenergies. 
Their vector components are given by 
%%%%%%%%%%%%%%%%%
\begin{align}
\Sigma^{\mu}_{p}  = &  g_{\omega N} \, \omega^{\mu} \, {\cal D}_{N}
+g_{\rho N}  \,  \rho^{\mu} \, {\cal D}_{N}~,
\label{Sigmav_p}\\
\Sigma^{\mu}_{n}  = &  g_{\omega N} \, \omega^{\mu} \, {\cal D}_{N}
- g_{\rho N}  \,  \rho^{\mu} \, {\cal D}_{N}~,
\label{Sigmav_n}
\end{align}
%%%%%%%%%%%%%%%%%
%%%%%%%%%%%%%%%%%
\begin{align}
\Sigma^{\mu}_{\Lambda}  = &  g_{\omega \Lambda} \, \omega^{\mu} \, {\cal D}_{\Lambda}~,
\label{Sigmav_l}
\end{align}
%%%%%%%%%%%%%%%%%
%%%%%%%%%%%%%%%%%
\begin{align}
\Sigma^{\mu}_{\Sigma^{+}}  = &  g_{\omega \Sigma} \, \omega^{\mu} \, {\cal D}_{\Sigma}
+g_{\rho \Sigma}  \,  \rho^{\mu} \, {\cal D}_{\Sigma}~,
\label{Sigmav_sp}\\
\Sigma^{\mu}_{\Sigma^{-}}  = &  g_{\omega \Sigma} \, \omega^{\mu} \, {\cal D}_{\Sigma}
-g_{\rho \Sigma}  \,  \rho^{\mu} \, {\cal D}_{\Sigma}~,
\label{Sigmav_sm}\\
\Sigma^{\mu}_{\Sigma^{0}}  = &  g_{\omega \Sigma} \, \omega^{\mu} \, {\cal D}_{\Sigma}~,
\label{Sigmav_s0}
\end{align}
%%%%%%%%%%%%%%%%%
%%%%%%%%%%%%%%%%%
\begin{align}
\Sigma^{\mu}_{\Xi^{-}}  = &  g_{\omega \Xi} \, \omega^{\mu} \, {\cal D}_{\Xi}
-g_{\rho \Xi}  \,  \rho^{\mu} \, {\cal D}_{\Xi}~,
\label{Sigmav_xim}\\
\Sigma^{\mu}_{\Xi^{0}}  = &  g_{\omega \Xi} \, \omega^{\mu} \, {\cal D}_{\Xi}
+g_{\rho \Xi}  \, \rho^{\mu} \, {\cal D}_{\Xi}
\label{Sigmav_xi0}
\,.
\end{align}
%%%%%%%%%%%%%%%%%
Similar expressions result for the scalar selfenergies. In the following the 
scalar and time-like component of the baryon selfenergy will be denoted as 
$S_{B}$ and $V_{B}$, respectively. Note that the selfenergies are explicitly 
momentum dependent due to the regulators $\nld_{B}=\nld_{B}(p)$ as specified 
below. 
The solutions of the Dirac equation are the 
standard Dirac-spinors with a proper normalization $N_{B}$
%%%%%%%%%%%%%%%%%
\begin{equation}
u_{B}(s,\pvec) = N_{B}
\left(
\begin{array}{c}
\varphi_{s} \\ \\
\ds \frac{ \sigmavec\cdot\pvec}{E^{*}_{B}+m^{*}_{B}}\varphi_{s}\\
\end{array}
\right)
\; , \label{Spinor}
\end{equation}
%%%%%%%%%%%%%%%%%
but now for quasi-free baryons $B$ with an in-medium energy 
%%%%%%%%%%%%%%%%%
\begin{equation}
E^{*}_{B} := E_{B} - V_{B}(p)~,
\label{onshell}
\end{equation}
%%%%%%%%%%%%%%%%%
and a Dirac mass
%%%%%%%%%%%%%%%%% 
\begin{equation}
m^{*}_{B} := m_{B} - S_{B}(p)~. 
\end{equation}
%%%%%%%%%%%%%%%%%
At a given momentum the single particle energy $E$ is 
obtained from the in-medium on-shell relation~(\ref{onshell}). These expressions 
are needed for evaluation of expectation values, for instance, the source terms 
of the meson-field equations. For the definition of the nuclear matter we need 
a conserved nucleon density. It is obtained from the time-like component of the 
Noether-current $J^{\mu}$ defined as
%%%%%%%%%%%%%%%%%
\begin{align}
J^{\mu} = \fac  \, \sum_{B=p,n} \, \pspace \, \frac{\Pi^{\mu}_{B}}{\Pi^{0}_{B}}
\label{current}
\end{align}
%%%%%%%%%%%%%%%%%
with the generalized $4$-momentum 
%%%%%%%%%%%%%%%%%
\begin{align}
\Pi^{\mu}_{B} = p^{*\mu}_{B}+ m^{*}_{B}
\Big(\partial_{p}^{\mu}S_{B} \Big)
- \Big(\partial_{p}^{\mu}\Sigma^{\beta}_{B} \Big) p^{*}_{B\beta}
\label{bigPi}
\end{align}
%%%%%%%%%%%%%%%%%
and the usual effective $4$-momentum 
%%%%%%%%%%%%%%%%%
\begin{equation}
p^{*\mu}_{B}=p^{\mu}-\Sigma^{\mu}_{B}
\,.
\end{equation}
%%%%%%%%%%%%%%%%%
The EoS (Equation of State) is obtained from the time-like components of the 
energy-momentum tensor. In nuclear matter the
resummation procedure of the NLD model results in the following expression 
%%%%%%%%%%%%%%%%%
\begin{align}
T^{\mu\nu} = 
\sum_{B} \fac \pspace \,
\frac{\Pi^{\mu}_{B} p^{\nu}}{\Pi^{0}_{B}} - g^{\mu\nu}\langle{\cal L}\rangle
\label{tensor}
\,,
\end{align}
%%%%%%%%%%%%%%%%%
from which the energy density $\varepsilon\equiv T^{00}$ and the pressure $P$ can
be calculated, see for details Ref.~\cite{nld}. 
Finally, the NLD meson-field equations in the RMF approach to nuclear matter 
can be resummed to the following forms
%%%%%%%%%%%%%%%%%
\begin{align}
m_{\sigma}^{2}\sigma + \frac{\partial U}{\partial\sigma} = & 
\sum_{B} g_{\sigma B} \,\Bla \overline{\psi}_{B}\nld_{B}\psi_{B}\Bra
= \sum_{B} g_{\sigma B} \, \rho_{sB}~, \\
m_{\omega}^{2}\omega = & 
\sum_{B} g_{\omega B} \,\Bla \overline{\Psi}_{B} \gamma^{0}\nld_{B}\Psi_{B}\Bra
= \sum_{B} g_{\omega B} \, \rho_{0B}
\label{mesonsNM}
\,,
\end{align}
%%%%%%%%%%%%%%%%%
with the scalar and vector density sources 
%%%%%%%%%%%%%%%%%
\begin{equation}
\rho_{sB} =  \fac \pspace \,
\frac{m^{*}_{B}}{\Pi^{0}_{B}} \, \nld_{B}(p)
~,
\label{dens_s}
\end{equation}
%%%%%%%%%%%%%%%%%
%%%%%%%%%%%%%%%%%
\begin{equation}
\rho_{0B} = \fac \pspace \, \frac{E^{*}_{B}}{\Pi^{0}_{B}} \, \nld_{B}(p)
\,.
\label{dens_0}
\end{equation}
%%%%%%%%%%%%%%%%%
The isovector densities are calculated through the standard isospin relations. 
For a hyperon with a given momentum relative to nuclear matter at rest (at a 
given nucleon density and isospin asymmetry) the 
mesonic sources contain only nucleons, that is $B=p,n$. 

The meson-field equations of motion  show a similar structure 
as those of the standard RMF approximation. However, the substantial difference 
between NLD and other conventional RMF models appears in the source terms which now contain 
in addition the momentum-dependent regulators $\nld_{B}$. This is an important 
feature of the NLD model. 
The cut-off leads naturally to a particular suppression of the vector field at 
high densities or high Fermi-momenta in agreement with phenomenology, as discussed 
in detail in the previous work~\cite{nld}. This feature 
is absent in conventional RHD approaches, except if one introduces by hand 
additional scalar/vector self-interactions. 

The key observable for general discussions related to momentum 
or energy dependencies of in-medium hadronic potentials is the Schroedinger-equivalent 
optical potential $U_{opt}$, which is a complex quantity. 
%%%%% changes in revision - 1
The imaginary part describes the scattering processes of a given particle, e.g., a hyperon, 
with a nucleon of the nuclear matter. The real part of the optical potential is related 
to the mean-field that a particle, e.g., a hyperon with a given momentum, experiences in 
the nuclear medium at a given density and isospin-asymmetry. The imaginary part of 
$U_{opt}$ cannot be calculated within a conventional RMF prescription. In RMF models one 
is usually interested in the real part of an optical potential that can be then 
examined in more realistic systems, for instance, in heavy-ion collisions or hadron-induced 
reactions within a relativistic transport theory. The missing imaginary part is then 
modelled within a collision term in terms of cross sections for elastic, quasi-elastic and 
inelastic channels with a proper counting of Pauli-Blocking effects. 

In the NLD model one cannot calculate precisely the imaginary part of $U_{opt}$. However, 
the NLD approach contains an explicit momentum dependence of the mean-fields, and thus, 
of the optical potential. This particular feature allow us to give, at least, estimations 
for the 
imaginary part of an optical potential too. This will be discussed in the case of the 
anti-hyperons, and we will mainly focus the study here on the real part of the optical 
potentials.

The real part of the Schroedinger-equivalent optical potential for hyperons is obtained 
from a non-relativistic reduction of the Dirac-equation and reads
\begin{align}
U_{opt}^{B} = -S_{B} + \frac{E_{B}}{m_{B}}V_{B} + 
\frac{1}{2m_{B}}\left( S_{B}^{2} - V_{B}^{2}\right)
\label{Uopt}
\,.
\end{align}
It describes the in-medium interaction of a baryon species $B$, e.g., a hyperon, 
with a momentum 
$p$ (or single-particle energy $E_{B}=E_{B}(p)$, see Eq.~(\ref{onshell})) 
relative to nuclear matter at rest at a given density and isospin asymmetry. 
We will use Eq.~(\ref{Uopt}) to compare the NLD results with the microscopic 
calculations from $\chi$-EFT and Lattice-QCD for the hyperon in-medium potentials. 

%%%%%%%%%%%%%%%% changes in revision - 1

%%%%%%%%%%%%%%%%%%%%%%%%%%%%%%%%%%%%%%%%%%%%%%%%%%%%%%%%%%%%%%%%%%%%%%%%%%%%%%%%%%%
\section{\label{sec3}Results and discussion}
%%%%%%%%%%%%%%%%%%%%%%%%%%%%%%%%%%%%%%%%%%%%%%%%%%%%%%%%%%%%%%%%%%%%%%%%%%%%%%%%%%%

\subsection{Nucleonic sector}

%%%%%%%%%%%%%%%
\begin{table}
\renewcommand*{\arraystretch}{1.4}
\begin{tabular}{|cccc|ccccccc|}
\hline\hline  
\multicolumn{4}{|c|}{\multirow{3}{*}{NLD parameters}} & $\Lambda_{sN}$ & $\Lambda_{vN}$ & $g_{\sigma N}$   & $g_{\omega N}$ & $g_{\rho N}$ & $b$ & $c$  
\\
& & & & $[\GeV]$ & $[\GeV]$ & & & & $[\frac{1}{fm}]$ & 
\\
\cline{5-11}
& & & & $0.95$ & $1.125~~$ & $~~10.08$ & $10.13$ & $3.50$ & $15.341$ & $-14.735~~$  \\
\hline\hline
\multicolumn{4}{|c|}{\multirow{3}{*}{Bulk saturation properties}} & 
$\rho_{sat}$  & $E_{b}$ & $K$ & $a_{sym}$ & & & 
\\
& & & & $[\frac{1}{fm^{3}}]$ & $[\frac{MeV}{A}]$ & $[\MeV]$ & $[\MeV]$ & & & 
\\
\cline{5-11}
& & & & $0.156$ & $-15.30$ & $251$ & $30$ & & & 
\\ 
\hline\hline
\end{tabular} 
\caption{
(Top) NLD parameters: meson-nucleon couplings 
$g_{mN},~(m=\sigma,\omega,\rho)$, $\sigma$ self-interaction constants $b,c$, and 
NLD cut-off for scalar ($\Lambda_{sN}$) and vector ($\Lambda_{vN}$) meson-nucleon isoscalar 
vertices. The isovector meson-nucleon cut-off is the same as the isoscalar-vector one. 
(Bottom) Bulk saturation properties of nuclear matter: saturation density 
$\rho_{sat}$, binding energy per nucleon $E_{b}$, compression modulus $K$ and asymmetry 
parameter $a_{sym}$ in the NLD model. See Ref.~\cite{nld} for more details. 
}
\label{tab1}
\end{table}
%%%%%%%%%%%%%%%

We briefly give the status of the NLD model for the in-medium nucleons, before 
starting the discussion on the in-medium hyperon potentials. As in detail 
discussed in~\cite{nld}, a momentum dependent monopole form 
\begin{equation}
\nld(p) = \frac{\Lambda^2}{\Lambda^2 + \vec{p\,}^{2}}
\label{monopole}
\end{equation}
for the regulators turned out to be very effective for a simultaneous description 
of the low and high 
density nuclear matter properties. An example is shown in table~\ref{tab1} for 
the extracted saturation properties together with the model parameters. It is seen 
that the NLD model leads to a very good description of the empirical values at 
saturation. The NLD EoS is rather soft and similar to the density dependence of 
Dirac-Brueckner-Hartree-Fock microscopic calculations. At high densities, however, 
the NLD EoS becomes 
stiff. This feature makes a prediction of the maximum mass of neutron stars of 
$2M_{\odot}$ possible even with a soft compression modulus. Note that the NLD model 
gives a  correct description of the Schroedinger-equivalent optical potential 
for in-medium protons and antiprotons simultaneously by imposing G-parity 
only~\cite{nld}. 

\subsection{Strangeness sector}

For the strangeness sector we consider again nuclear matter at rest, at a given 
density, isospin-asymmetry and at zero temperature, 
in which hyperons ($\Lambda, \Sigma, \Xi$) are situated at a given 
momentum relative to the nuclear matter at rest. The quantity of interest 
will be the optical potential $U_{opt}$ of the in-medium hyperons , 
see Eq.~(\ref{Uopt}). 
Since there is no experimental information on the momentum dependence of the 
in-medium hyperonic potentials, we use for our comparisons the recent microscopic 
calculations from Refs.~\cite{eft} (see also Ref.~\cite{eft-new}) 
and~\cite{LQCD} as a guidance. They are based on the 
$\chi$-EFT approach in Next-To-Leading (NLO) order and to Lattice-QCD. 

In the NLD calculations we assume for the in-medium hyperon interactions 
no additional parameters except of the strangeness cut-off of the 
hyperons. That is, 
the various hyperon-nucleon couplings are fixed from the corresponding 
nucleon-nucleon ones by means of SU(6). The hyperon cut-offs retain 
their monopole form as in Eq.~(\ref{monopole}). In particular, they take the 
form
%%%%%%%%%%%%%%%%
\begin{equation}
\nld_{Y}(p) = 
\frac{\Lambda^{2}_{\gamma_{1}}}{\Lambda^{2}_{\gamma_{2}} + \vec{p\,}^{2}}\,,
\label{monopole2}
\end{equation}
%%%%%%%%%%%%%%%%
with $\gamma=\sigma,~\omega,~\rho$ indicating the cut-off values for the 
hyperon-nucleon $\sigma,~\omega$- and $\rho$-vertices, respectively, and 
$Y=\Lambda, \Sigma, \Xi$ denotes the hyperon type. In principle, one could use a 
single cut-off $\Lambda_{\gamma_{1}}=\Lambda_{\gamma_{2}}=\Lambda_{\gamma}$ 
for each meson-hyperon vertex. However, in order to describe the 
non-trivial momentum dependence of the microscopic calculations as precise 
as possible we allow for different cut-off values for the vector-isoscalar 
$\omega$- and vector-isovector $\rho$-hyperon vertices, 
as shown in Eq.~(\ref{monopole2}). For the isoscalar meson-hyperon interactions 
a single cut-off $\Lambda_{\sigma}=\Lambda_{\sigma_{1}}=\Lambda_{\sigma_{2}}$ 
for each hyperon type is used. This prescription was found to be the most appropriate 
one when comparing to the microscopic calculations. In fact, the scalar-like  
interactions are in any case better controlled with increasing density (respectively 
momentum) by $m^{\star}/E^{\star}$-suppression factors while the vector-like 
vertices do not include them, besides the NLD-regulators in the source terms of 
the meson-field equations (\ref{dens_s},\ref{dens_0}). Note that 
$\Pi^{0}=E^{\star}$ for momentum-dependent regulators $\Pi^{0}=E^{\star}$ 
and for each baryon type B. Similar studies concerning the peculiar role of the 
vector $\omega$-meson exist in the literature. For instance, in 
Refs.~\cite{nl-omega-1,nl-omega-2,nl-omega-3} 
non-linear quadratic $\omega$-field contributions were considered as an alternative 
approach for the vector-like interaction Lagrangian leading to more complex density 
dependencies of their mean-fields. In the NLD model all higher-order 
non-linear terms are summed up into regulators. The novel feature of NLD is 
that these regulators mediate a non-linear density and, at the same time, a 
non-linear momentum dependence of in-medium potentials not only for nucleons, but 
for hyperons too. This will become clear in the following discussions.

%%%%%%%%%%%%%%%%%%
\begin{figure}[t]
\begin{center}
\unitlength1cm
\begin{picture}(12.,10.0)
\put(-1.25,0.0){
\makebox{\includegraphics[clip=true,width=0.9\columnwidth,angle=0.]
		 {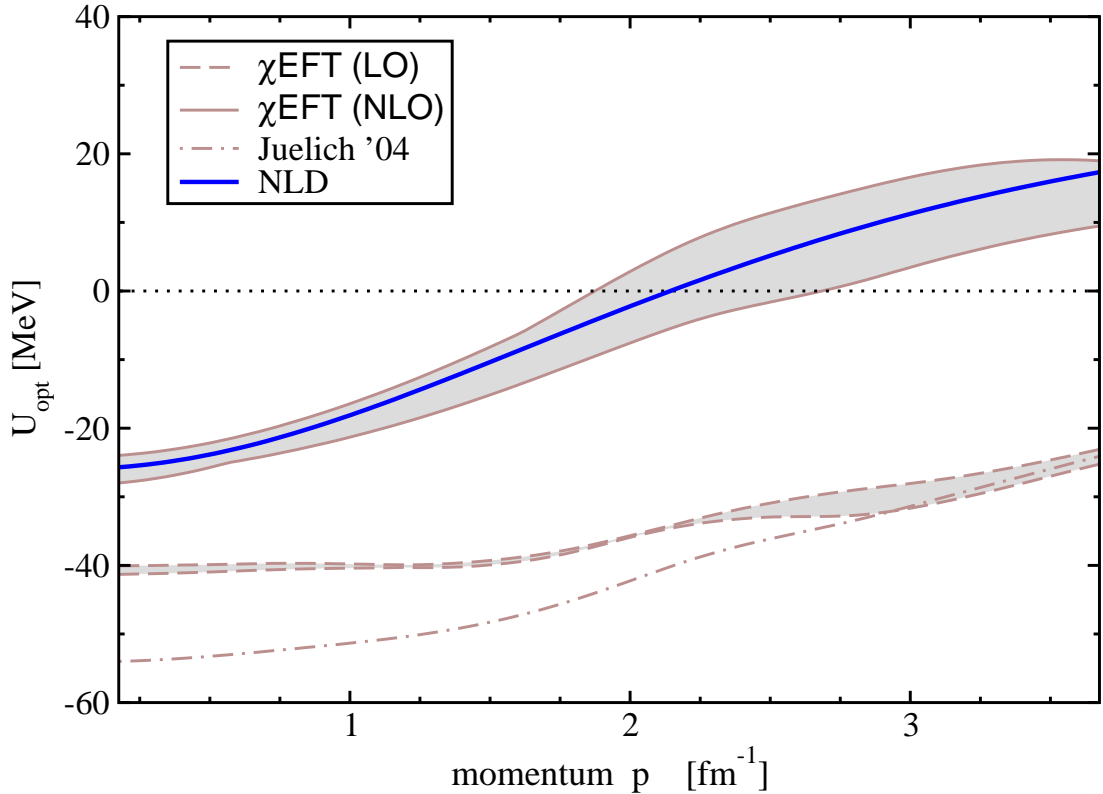}
		 }
}
\end{picture}
\caption{Optical potential of $\Lambda$-hyperons as function of their momentum $p$ 
in symmetric nuclear matter at saturation density. The NLD-results (thick-solid curve) 
are compared with $\chi$-EFT microscopic calculations (taken from~~\cite{eft}) 
at different orders 
LO (band with closed dashed borders) and NLO (band with closed solid borders)~\cite{eft}. 
Further microscopic calculations from the J\"{u}lich group (dot-dashed curve) are 
shown too~\cite{juelich}.
}
\label{Fig1}
\end{center}
\end{figure}
\begin{figure}[t]
\begin{center}
\unitlength1cm
\begin{picture}(12.,10.0)
\put(-1.25,0.0){
\makebox{\includegraphics[clip=true,width=0.9\columnwidth,angle=0.]
		 {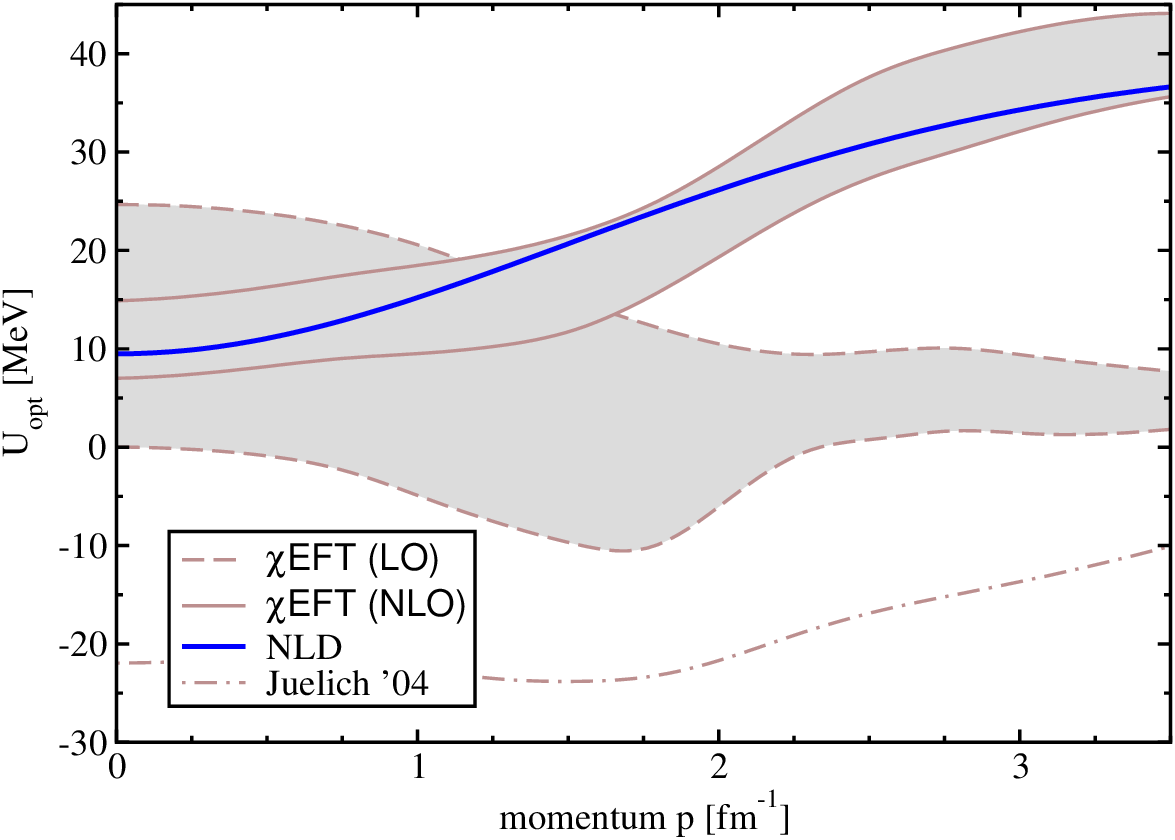}
		 }
}
\end{picture}
\caption{Same as in Fig.~\ref{Fig1}, but for $\Sigma$-hyperon.
}
\label{Fig2}
\end{center}
\end{figure}

%%%%%%%%%%%%%%%%%%
At first, the cut-offs of the hyperons have to be determined. 
The strangeness-$S=1$ cut-offs are adjusted to the corresponding hyperonic 
optical potentials at saturation density of symmetric and cold nuclear matter 
from $\chi$-EFT calculations. 
This is shown in Fig.~\ref{Fig1} for the optical potential of $\Lambda$-hyperons. 
The gray bands correspond to the microscopic calculations at different orders in 
$\chi$-EFT, while the solid curve represents the NLD result. 
At low momenta the $\Lambda$ in-medium interaction is attractive, but it becomes 
repulsive at high momenta. The non-trivial momentum dependence in NLD arises from the 
explicitly momentum dependent regulators which show up twice: in 
the scalar and vector selfenergies and in the source terms of the meson fields. 
As a consequence, the cut-off regulates the $\Lambda$-potential not only 
at zero momentum, but particularly over a wide momentum region. The 
in-medium $\Lambda$-potential does not diverge with increasing $p$-values (not shown here), 
but it saturates. Furthermore, the in-medium $\Lambda$-potential at zero kinetic 
energy leads to a value of $U_{opt}^{\Lambda}\simeq -28~\MeV$, which is consistent 
with the NLO-calculations and also consistent with phenomenology. Therefore  
it exists an appropriate choice of cut-off regulators that do reproduce the 
microscopic calculations over a wide momentum range up to $p\simeq 1~\GeV$ very well.
A similar picture occurs for the in-medium potential of $\Sigma$-hyperons, as shown 
in Fig.~\ref{Fig2}. The NLD cut-off for the $\Sigma$-particles can be regulated in 
such way to reproduce a repulsive potential at vanishing momentum with a weak momentum 
dependence at finite $\Sigma$-momentum. Again, the NLD calculations are able to 
describe the microscopic $\chi$-EFT results in NLO very well. The corresponding values 
for the strangeness cut-offs are tabulated in~\ref{tab2}. Even if the origin of 
the cut-offs is different between the NLD model and the microscopic calculations, 
it may be interesting to note that these NLD cut-off values are close to the region 
between 500 and 650 \GeV~used in the $\chi$-EFT calculations. 

%%%%%%%%%%%%%%%
\begin{table}
\renewcommand*{\arraystretch}{1.4}
\begin{tabular}{|c c c c c|c c c c c|c c c c c|}
\hline\hline  
\multicolumn{5}{|c|}{$\Lambda$ cut-off} & \multicolumn{5}{|c|}{$\Sigma$ cut-off} & 
\multicolumn{5}{|c|}{$\Xi$ cut-off}
\\
\hline
$\Lambda_{\sigma}$ & $\Lambda_{\omega_1}$ & $\Lambda_{\omega_2}$ & 
$\Lambda_{\rho_1}$ & $\Lambda_{\rho_2}$ & 
$\Lambda_{\sigma}$ & $\Lambda_{\omega_1}$ & $\Lambda_{\omega_2}$ & 
$\Lambda_{\rho_1}$ & $\Lambda_{\rho_2}$ & 
$\Lambda_{\sigma}$ & $\Lambda_{\omega_1}$ & $\Lambda_{\omega_2}$ & 
$\Lambda_{\rho_1}$ & $\Lambda_{\rho_2}$
\\
0.7 & 0.85 & 0.79 & -- & -- & 
0.67 & 0.95 & 0.79 & 0.47 & 0.47 & 
0.6 & 0.8 & 0.71 & 1.3 & 1.2
\\ 
& & & & & & & & 0.63 & 0.5 & & & & &
\\
\hline\hline
\end{tabular} 
\caption{
$\Lambda$, $\Sigma$ and $\Xi$ cut-offs for $\sigma$- ($\Lambda_{\sigma}$), 
$\omega$- ($\Lambda_{\omega_{1,2}}$) and $\rho$-hyperon-nucleon 
($\Lambda_{\rho_{1,2}}$) 
vertices in units of \GeV. In the cases for $\Sigma$ and $\Xi$ the isospin 
cut-offs ($\Lambda_{r1,2}$) are relevant for the charged particles only. 
For the $\Sigma$-hyperon different cut-off values $\Lambda_{\rho_{1,2}}$ 
are used for $\Sigma^{-}$ (upper line) and for $\Sigma^{+}$ (bottom line).
}
\label{tab2}
\end{table}
%%%%%%%%%%%%%%%

%%%%%%%%%%%%%%%%%%
%%%%%%%%%%%%%%%%%%
We emphasize again the non-trivial momentum dependence of the in-medium 
hyperon-potentials, as manifested in the $\chi$-EFT calculations at different 
orders, see for instance Ref.~\cite{eft}.
%As reported in Ref.~\cite{eft}, the microscopic calculations 
%become more robust against parameter variations with increasing leading order 
%in the underlying interaction kernel. 
This prescription modifies the momentum 
dependencies in such a complex way, which cannot be reproduced in standard RMF 
models by imposing SU(6) arguments. Furthermore, 
any standard RMF model leads to a divergent behaviour of optical potentials at high 
momenta. Note that a weak repulsive character of the $\Sigma$-potential, as proposed 
by the microscopic calculations, cannot be 
achieved in conventional RMF. The momentum-dependent NLD model resolves these issues 
effectively through momentum cut-offs of natural hadronic scale. Since we are dealing 
with hadronic matter, values of hadronic scale in the \GeV-regime for the NLD regulators 
seem to be an adequate choice. 

\begin{figure}[t]
\begin{center}
\unitlength1cm
\begin{picture}(12.,8.0)
\put(-1.25,0.0){
\makebox{\includegraphics[clip=true,width=0.9\columnwidth,angle=0.]
		 {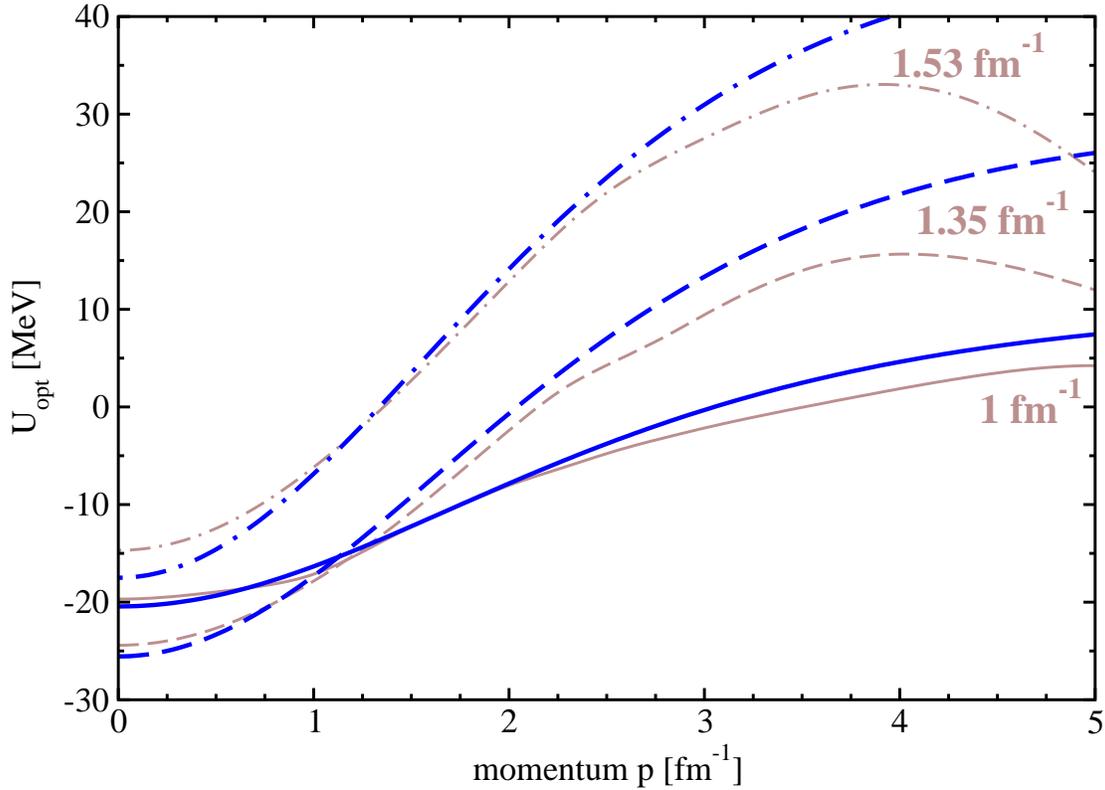}
		 }
}
\end{picture}
\caption{Optical potential of $\Lambda$-hyperons versus their momentum at various 
densities of symmetric nuclear matter, as indicated by the Fermi-momenta 
in units of 1/fm${}^{-1}$. The NLD calculations at these three Fermi-momenta 
(thick-solid, thick-dashed and thick-dot-dashed curves) are compared 
to the $\chi$-EFT calculations at NLO~\cite{eft}.
}
\label{Fig3}
\end{center}
\end{figure}

\begin{figure}[t]
\begin{center}
\unitlength1cm
\begin{picture}(12.,8.0)
\put(-1.25,0.0){
\makebox{\includegraphics[clip=true,width=0.9\columnwidth,angle=0.]
		 {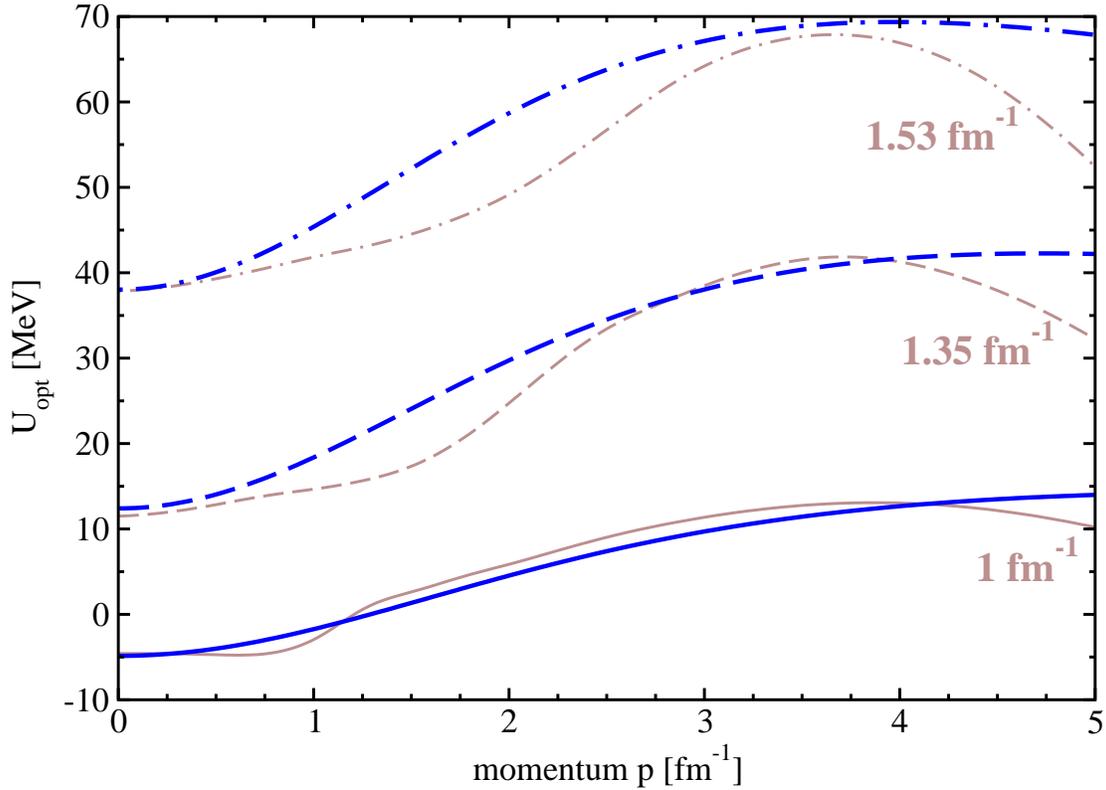}
		 }
}
\end{picture}
\caption{Same as in Fig.~\ref{Fig3}, but for the $\Sigma$-hyperons.
}
\label{Fig4}
\end{center}
\end{figure}

So far we have discussed the momentum dependence of the $\Lambda$ and $\Sigma$ hyperons 
at saturation density (Figs.~\ref{Fig1} and \ref{Fig2}). These comparisons served also 
as a guideline for the NLD cut-offs for the $\Lambda$ and $\Sigma$ baryons. Now we 
discuss the predictive power of the NLD approach by comparing in more detail the 
density and momentum dependence of the NLD formalism with the microscopic 
$\chi$-EFT calculations. 
This is shown in Figs.~\ref{Fig3} and \ref{Fig4}, where the momentum dependence of 
the $\Lambda$ (Fig.~\ref{Fig3}) and $\Sigma$ (Fig.~\ref{Fig4}) particles is displayed 
again, but now at various densities of symmetric nuclear matter. At first, the 
$\Lambda$ and $\Sigma$ optical potentials become more repulsive with increasing 
nuclear matter density in NLD. However, the non-trivial momentum and density dependence, 
as manifested in the NLD selfenergies and the meson-field sources, weakens the in-medium 
potentials with increasing momentum. In particular, the NLD model predicts 
astonishingly well the complex microscopic behaviours in momentum and at various 
densities of symmetric nuclear matter. 

In asymmetric matter besides the standard iso-scalar and iso-vector vertices ($\sigma$ 
and $\omega$ meson fields, respectively) the iso-vector and Lorentz-vector $\rho$-meson 
must be taken into account. In NLD we assume again a monopole form for the $\rho$-meson 
coupling to the hyperons too by using the coupling constant of table~\ref{tab1} 
and the cut-off values of table~\ref{tab2} for the isospin sector. Relevant are the 
cut-off values $\Lambda_{\rho_{1,2}}$ for the charged $\Sigma^{\pm}$-hyperons. They 
have been fixed from the corresponding $\chi$-EFT calculations for $\Sigma^{-}$ and 
$\Sigma^{+}$ at saturation density. The NLD calculations for the neutral 
$\Lambda$- and $\Sigma^{0}$-hyperons are free of parameters here. 

\begin{figure}[t]
\begin{center}
\unitlength1cm
\begin{picture}(12.,8.0)
\put(-1.25,0.0){
\makebox{\includegraphics[clip=true,width=0.9\columnwidth,angle=0.]
		 {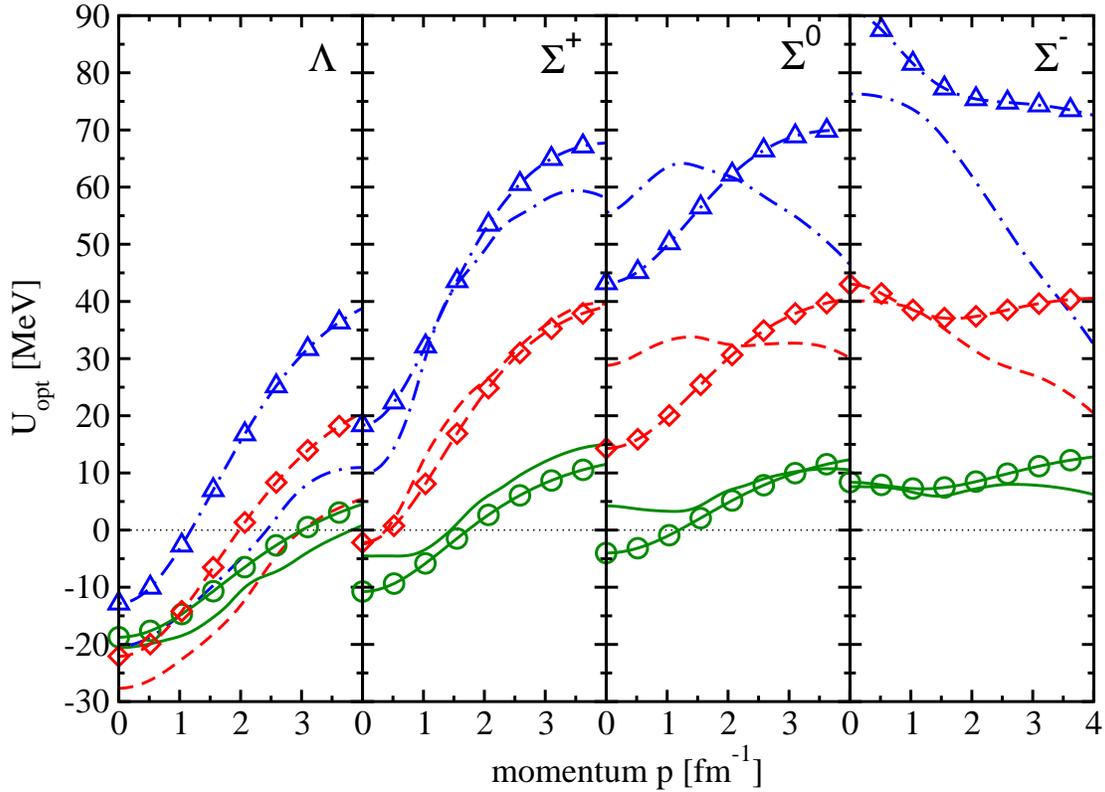}
		 }
}
\end{picture}
\caption{
Optical potentials for hyperons (as indicated) versus their momentum 
for pure neutron matter. 
Solid curves with symbols indicate the NLD calculations while pure 
curves without symbols are the microscopic $\chi$-EFT results at NLO 
from Ref.~\cite{eft}. 
Green pairs (circles-solid for NLD and solid for $\chi$-EFT) refer to 
low density of $p_{F}=1$ fm${}^{-1}$, 
red pairs (diamonds-dashed for NLD and dashed for $\chi$-EFT) refer to 
saturation density of $p_{F}=1.35$ fm${}^{-1}$ and 
blue pairs (triangles-dot-dashed for NLD and dot-dashed for $\chi$-EFT) refer to 
a density of $p_{F}=1.53$ fm${}^{-1}$.
}
\label{Fig5a}
\end{center}
\end{figure}

The results for pure neutron matter at three different baryon densities are summarized 
in Fig.~\ref{Fig5a}. The NLD model does predict the general microscopic trends. 
In particular, in the case of the neutral hyperons ($\Lambda$ and $\Sigma^{0}$), 
where within the RMF approximation the $\rho$-meson does not appear at all, one 
would expect identical results between symmetric and pure neutron matter 
(at same total baryon density and momentum). This is in fact not the case. 
There is an inherent isospin dependence in the source terms of the meson-field equations, 
see Eqs.~(\ref{mesonsNM}) even for the $\sigma$- and $\omega$-fields. 
The upper limits in those integrals~(\ref{dens_s},~\ref{dens_0}) are different for 
protons and neutrons between symmetric and asymmetric nuclear matter at the same total 
density. This leads to a different cut value in the regulators $\nld_{p,n}$ and thus to 
a different result between symmetric and asymmetric matter. This NLD feature induces a 
hidden isospin dependence which is qualitatively consistent with the microscopic 
calculations at the three total densities as indicated in Fig.~\ref{Fig5a} for the 
"isospin-blind" hyperons. Concerning the charged $\Sigma^{\pm}$-hyperons, 
the comparison between NLD and $\chi$-EFT calculations is obviously at best for 
densities close to saturation. In general, the NLD predictions follow 
satisfactorily the details of the microscopic in-medium potentials as function of 
momentum and matter density. 

\begin{figure}[t]
\begin{center}
\unitlength1cm
\begin{picture}(12.,10.0)
\put(-1.25,0.0){
\makebox{\includegraphics[clip=true,width=0.9\columnwidth,angle=0.]
		 {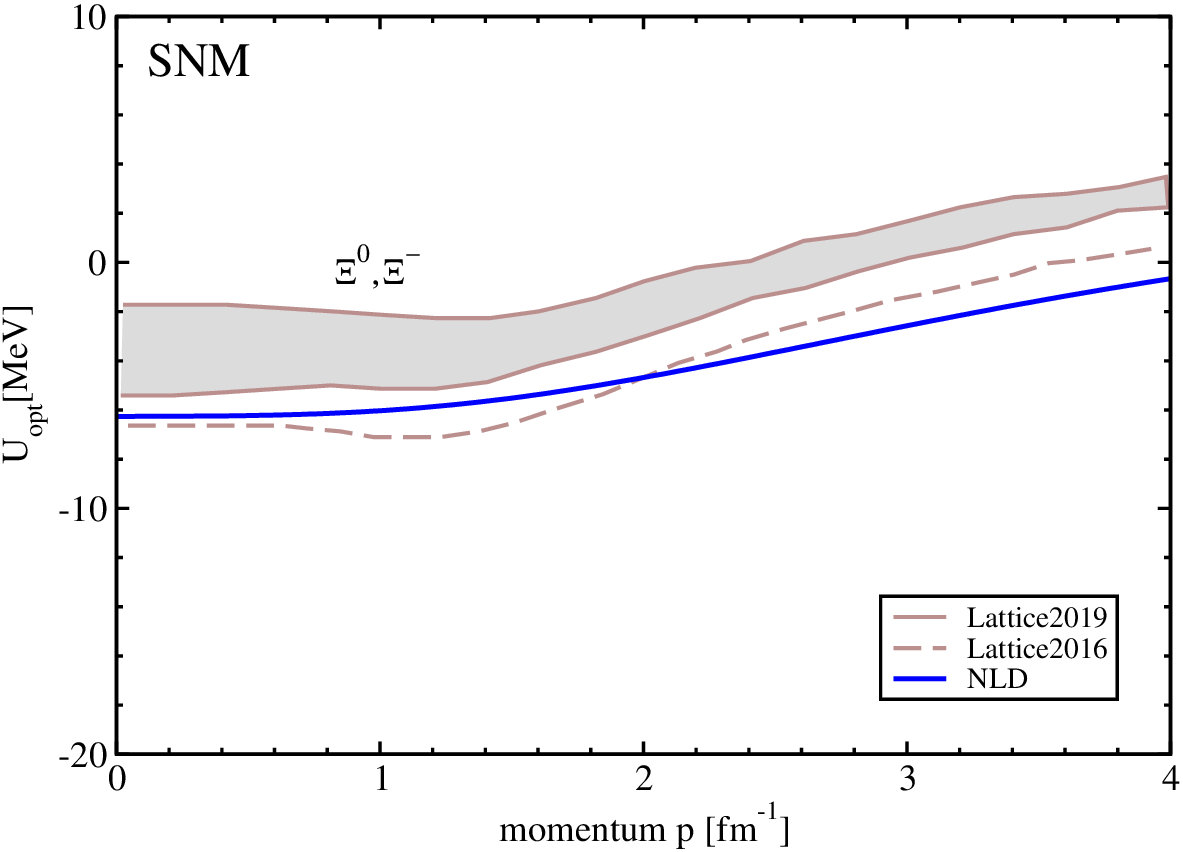}
		 }
}
\end{picture}
\caption{Optical potential of cascade hyperons versus their momentum for symmetric 
nuclear matter (SNM) at saturation density. The solid curve indicates the 
NLD predictions while the dashed curve and the gray band refer to recent 
Lattice calculations from Refs.~\cite{LQCD} (Lattice2016 and Lattice2019). 
}
\label{Fig7}
\end{center}
\end{figure}

\begin{figure}[t]
\begin{center}
\unitlength1cm
\begin{picture}(12.,10.0)
\put(-1.25,0.0){
\makebox{\includegraphics[clip=true,width=0.9\columnwidth,angle=0.]
		 {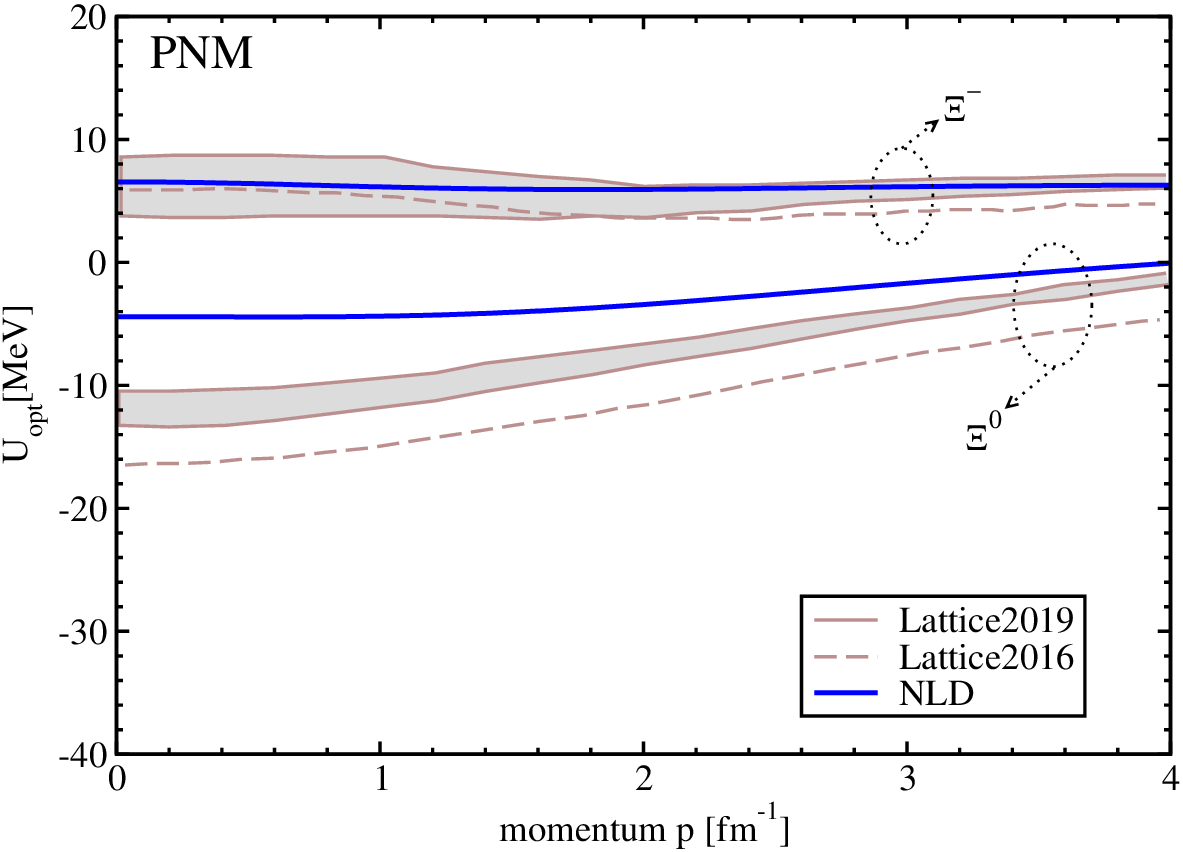}
		 }
}
\end{picture}
\caption{Same as in Fig.~\ref{Fig7}, but for pure neutron matter (PNM). 
The curves and bands belonging to $\Xi^{-}$ and $\Xi^{0}$ are indicated in this figure. 
}
\label{Fig8}
\end{center}
\end{figure}

Finally we discuss the in-medium properties of the cascade-hyperons as shown in 
Figs.~\ref{Fig7} and \ref{Fig8} for symmetric nuclear matter (SNM) and pure neutron 
matter (PNM). Here we apply for comparison the latest microscopic calculations from 
Lattice-QCD. The same NLD scheme with appropriate monopole-type regulators 
leads to the results in Fig.~\ref{Fig7} for symmetric nuclear matter at saturation density. 
It is seen that a simple monopole-like regulator with hadronic cut-off values 
can explain the microscopic Lattice calculations. Indeed, a soft attractive potential 
for in-medium $\Xi$-hyperons is obtained in the NLD model over a wide momentum range. 
The prediction of NLD is then displayed in Fig.~\ref{Fig8} 
for pure neutron matter but at the same total density at saturation as in the previous 
figure. The hidden isospin-dependence modifies slightly the momentum dependence of the 
neutral $\Xi^{0}$-hyperon. In this case the Lattice calculations are reproduced only 
qualitatively by the NLD model, while for the charged cascade partner ($\Xi^{-}$) the 
comparison between NLD and Lattice is very well for pure neutron matter at saturation and 
over a broad region in single-particle cascade-momentum. 

\begin{figure}[t]
\begin{center}
\unitlength1cm
\begin{picture}(12.,10.0)
\put(-1.25,0.0){
\makebox{\includegraphics[clip=true,width=0.9\columnwidth,angle=0.]
		 {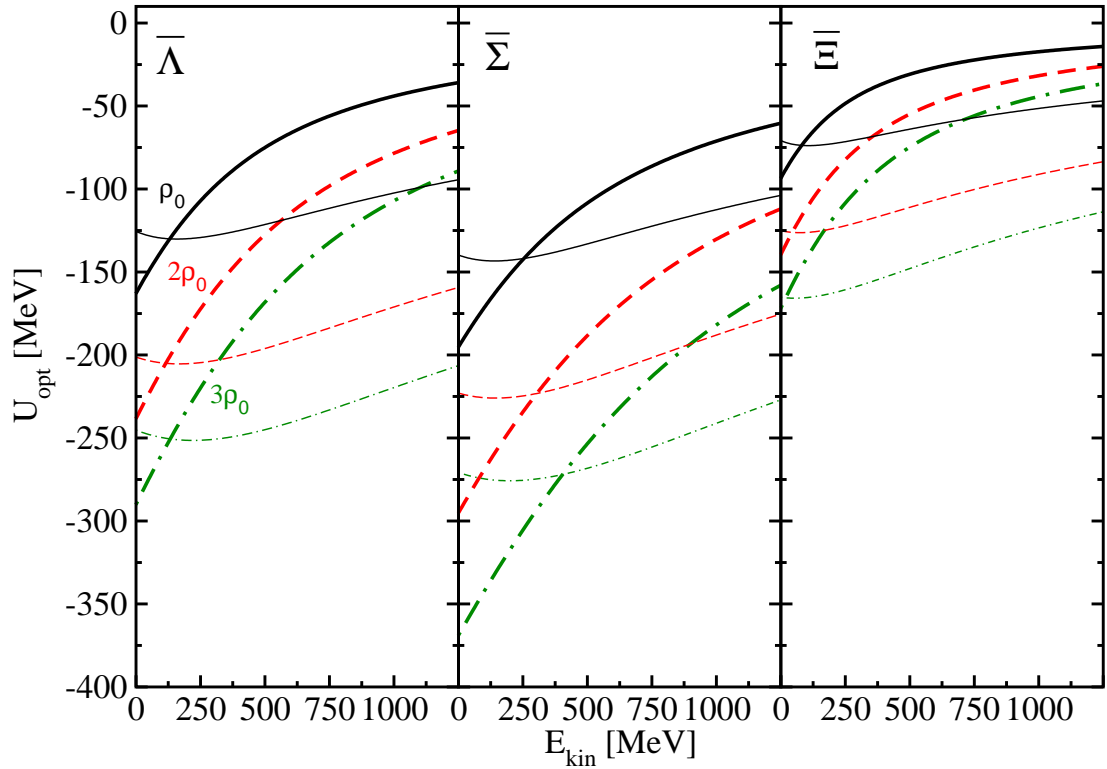}
		 }
}
\end{picture}
\caption{
Optical potentials for anti-hyperons versus their kinetic energy for symmetric nuclear 
matter (SNM) at various densities, as indicated. The NLD predictions for saturation 
density $\rho_{0}$ (thick-solid) and higher densities of $2\rho_{0}$ (thick-dashed) and 
$3\rho_{0}$ (thick-dashed-dot) are shown. For the anti-hyperons we show estimates for 
the imaginary part of their optical potentials too at saturation density $\rho_{0}$ 
(thin-solid), at $2\rho_{0}$ (thin-dashed) and at $3\rho_{0}$ (thin-dashed-dot). 
}
\label{Fig_AntiHyp}
\end{center}
\end{figure}

In the future experiments such as those at FAIR the in-medium properties of 
anti-hadrons will be investigated too. We thus give predictions for anti-hyperon 
in-medium potentials too. We recall the novel feature of the NLD formalism~\cite{nld}, 
that is, a parameter free predictions for anti-baryon optical potentials in the 
spirit of G-parity. In fact, once the cut-off parameters are fixed from saturation 
properties, the application of NLD to anti-matter gave very successful results by 
imposing G-Parity only. Note that in conventional RMF models one has to introduce by 
hand additional scaling factors in order to reproduce the weak attractiveness of the 
anti-proton optical potential at vanishing momenta~\cite{Larionov}. 
We therefore use the same NLD formalism for the description of anti-hyperons too and 
performed additional calculations for the $\overline{\Lambda}$, $\overline{\Sigma}$ and 
$\overline{\Xi}$ optical potentials as function of momentum and density. These results are 
shown in Fig.~\ref{Fig_AntiHyp} for anti-$\Lambda$ (left), anti-$\Sigma$ (middle) and 
anti-$\Xi$ optical potentials versus their momentum at three densities of symmetric 
nuclear matter. 
Due to the negative sign in the Lorentz-vector component of the hyperon self-energy these 
potentials are in general attractive over a wide momentum range. Compared to the anti-proton 
potential at saturation these potentials are less attractive with a similar dependence on 
single-particle momentum. 

%%%%%%%%%%%%%%%%%%%%%%%%%%% changes in revision -2 
Since for anti-hyperons we make predictions and for anti-particles in general one may expect 
significant contributions to the imaginary part of $U_{opt}$ too, we briefly discuss 
the imaginary part of the anti-hyperon optical potentials too. An exact treatment of 
the imaginary part of the optical potential is not possible within an RMF model. However, 
within the NLD approach one can estimate the strength of $Im~U_{opt}$ from dispersion 
relations~\cite{nld}. 
%This is possible in the NLD model because of the explicit momentum 
%dependence of the mean-fields, and thus, due to an explicit momentum dependence of the 
%real part of the optical potential. As discussed in Ref.~\cite{nld}, 
%the imaginary part of the Schr\"{o}dinger 
%equivalent optical potential is given by 
%%%%%%%%%%%%%%%%%
%\begin{equation}
%Im~U_{\rm opt}(p) = -\frac{2p}{\pi} \; {\cal P} \int_{0}^{\infty} 
% \frac{ Re~U_{\rm opt}(p^{\prime \;})}{p^{\prime\;2}-p^{2}} dp^{\prime}
%\:.
%\end{equation}
%%%%%%%%%%%%%%%%%
%Here $p\equiv |\vec{p}\;|$ stands for the particle momentum and 
%${\cal P}$ denotes the principal value. We note that 
%this equation makes sense only if $Re~U_{\rm opt}$ 
%does not diverge as function of the particle momentum. This property of 
%$Re~U_{\rm opt}$ is realized in the NLD approach. 
This prescription was successfully applied to the antiproton case 
in a previous work (see Ref.~\cite{nld}), thus we apply it here for the anti-hyperons too. 
The results for $Im~U_{opt}$ are shown in the same figure~\ref{Fig_AntiHyp} by the 
thin curves. One generally observes a strong contribution to the in-medium 
anti-hyperon interactions from the imaginary parts of the optical potentials too. 
These contributions are quite similar to the imaginary potential of antiprotons with a 
value around -150~\MeV~at very low kinetic energies (see for instance in~\cite{nld} the 
second citation of 2015). However, in the antiproton-case the imaginary potential is 
rather strong relative to its real part, while for anti-hyperons both parts of the potential 
are sizeable. 
Even if the NLD results for the $Im~U_{opt}$ are only estimations, we can give a 
physical interpretation. In antinucleon-nucleon scattering annihilation can occur through 
the production of light pions. On the other hand, the interaction of anti-hyperons with 
nucleons can happen via the production of the heavier kaons due to strangeness conservation, 
which may influence the imaginary potential at low energies. 
This might be one reason why the imaginary part of the anti-hyperon optical potential 
is comparable with its corresponding real part particularly at very low energies. 
These calculations can be applied to anti-hadron induced reactions in the spirit of 
relativistic transport theory and can be tested in the future experiments at FAIR. 
%%%%%%%%%%%%%%%%%%%%%%%%%%% changes in revision -2

%%%%%%%%%%%%%%%%%%
%%%%%%%%%%%%%%%%%%

%%%%%%%%%%%%%%%%%%
%%%%%%%%%%%%%%%%%%

%%%%%%%%%%%%%%%%%%% 5 
%%%%%%%%%%%%%%%%%%%

%%%%%%%%%%%%%%%%%% 6
%%%%%%%%%%%%%%%%%%

%%%%%%%%%%%%%%%%%% 7 
%%%%%%%%%%%%%%%%%%

%%%%%%%%%%%%%%%%%%%%%%%%%%% changes in revision -1 
%%%%%%%%%%%%%%%%%% 8 
%%%%%%%%%%%%%%%%%%
%%%%%%%%%%%%%%%%%%%%%%%%%%% changes in revision -1 

%%%%%%%%%%%%%%%%%%%%%%%%%%%%%%%%%%%%%%%%%%%%%%%%%%%%%%%%%%%%%%%%%%%%%%%%%%%%%%
\section{Summary}
%%%%%%%%%%%%%%%%%%%%%%%%%%%%%%%%%%%%%%%%%%%%%%%%%%%%%%%%%%%%%%%%%%%%%%%%%%%%%%

We have investigated the properties of strangeness particles inside nuclear matter in 
the framework of the NLD approach. The NLD model is based on the simplicity of the 
relativistic mean-field theory, but it includes the missing momentum dependence in 
a manifestly covariant fashion. This is realized by the introduction of non-linear 
derivative series in the interaction Lagrangian. In momentum space this prescription 
leads to momentum dependent regulators, which are determined by a cut-off. The NLD 
approach does not only resolve the optical potential issues of protons and antiprotons 
at high momenta, but 
it affects the density dependence. That is, the cut-off regulators make the EoS 
softer at densities close to saturation and stiffer at very high densities relevant 
for neutron stars. 

Because of the successful application of the NLD model to infinite nuclear matter (and 
to finite nuclei~\cite{nldnuclei}), it is a natural desire to extend this approach to 
hadronic matter by taking strangeness degrees of freedom into account. This is 
realized in the spirit of SU(6) symmetry. We applied the NLD model to the description 
of in-medium hyperon interactions for ordinary nuclear matter. 
It was found that 
the strangeness cut-off regulates the momentum dependence of the optical 
potentials of hyperons in multiple ways. At first, the optical potentials do not 
diverge with increasing hyperon momentum. Furthermore, the NLD model predicts an 
attractive $\Lambda$-optical potential at low momenta, which becomes repulsive 
at high energies and finally saturates. In particular, it is possible 
to predict a weak and repulsive in-medium interaction for $\Sigma$-hyperons 
inside nuclear matter at saturation density. These results are in consistent 
agreement with calculations based on the chiral effective field theory. 
Regarding $\Xi$-hyperons, the NLD predictions turned out to be in agreement with recent 
Lattice-QCD calculations. In symmetric nuclear matter the cascade 
optical potential is attractive and it follows the Lattice-QCD results. In 
pure neutron matter the isospin-separation as predicted by the NLD model 
agrees with the Lattice-QCD behaviours qualitatively. While the potential of 
the neutral cascade particle remains attractive, the $\Xi^{-}$-hyperon 
shows a weak repulsion in neutron matter. The weak repulsion of those hyperons 
may likely effect to a stiffer EoS for neutron star matter. 

%%%%%%%%%%%%%%%%%%%%%%%%%%% changes in revision -3
We briefly discussed the imaginary part of $U_{opt}$ of anti-hyperons too. These 
estimations indicate a significant contribution of the imaginary part to the 
anti-hyperon dynamics that could be explored in anti-hadron induced reactions. 
For instance, the present calculations can be tested in anti-proton induced 
reactions and in reactions with secondary $\Xi$-beams, as they are planned at FAIR 
in the future PANDA experiment. 
%%%%%%%%%%%%%%%%%%%%%%%%%%% changes in revision -3 

Obviously this study is relevant not only for hadron physics, but also for nuclear 
astrophysics. The application of the NLD approach to $\beta$-equilibrated 
compressed matter is under progress, in order to investigate the hyperon-puzzle 
in neutron stars. Another interesting application concerns the dynamics of 
neutron star binaries. To do so, an extension to hot and compressed hadronic 
matter is necessary and under progress too. Note that the NLD formalism is 
fully thermodynamically consistent, which is an important requirement before 
applying it to hot and dense systems. In summary, we conclude the relevance of 
our studies for future experiments at FAIR and for nuclear astrophysics. 

%%%%%%%%%%%%%%%%%%%%%%%%%%%%%%%%%%%%%%%%%%%%%%%%%%%%%%%%%%%%%%%%%%%%%%%%%%%%%%%%
%%%%%%%%%%%%%%%%%%%%%%%%%%%%%%%%%%%%%%%%%%%%%%%%%%%%%%%%%%%%%%%%%%%%%%%%%%%%%%%%
\section*{Acknowledgments}
This work is partially supported by COST (THOR, CA 15213) and by the European 
Union's Horizon 2020 research and innovation programme under grant agreement 
No. 824093. We also acknowledge H.~Lenske and J.~Haidenbauer for fruitful 
discussions and for providing us the $\chi$-EFT calculations. 

%%%%%%%%%%%%%%%%%%%%%%%%%%%%%%%%%%%%%%%%%%%%%%%%%%%%%%%%%%%%%%%%%%%%%%%%%%%%%%%%
%%%%%%%%%%%%%%%%%%%%%%%%%%%%%%%%%%%%%%%%%%%%%%%%%%%%%%%%%%%%%%%%%%%%%%%%%%%%%%%%

%\bibliographystyle{h-physrev}	% (uses file "plain.bst")
%\bibliographystyle{apsrev4-1}	% (uses file "plain.bst")
%\bibliography{literature}		% expects file "myrefs.bib"

\section*{References}

\begin{thebibliography}{10}

\bibitem{NS1}
P.~Demorest, T.~Pennucci, S.~Ransom, M.~Roberts, and J.~Hessels,
\textit{Nature} {\bf 467} (2010) 1081.

\bibitem{NS2}
J.~Antoniadis {\it et al.}, Science {\bf 340} (2013) 6131.

\bibitem{NSnew}
M. Linares, T. Shahbaz and J. Casares, 
\textit{Astrophys. J.} \textbf{859 (1)} (2018) 54\\
H.T. Cromartie, E. Fonseca, et al., 
\textit{Nature Astronomy} \textbf{4} (2020) 72.

\bibitem{Lattimer14}
J.M.~Lattimer, A.W.~Steiner, 
\textit{Astrophys.~J.} {\bf 784} (2014) 123.

\bibitem{softeos1}
  T.~Kl\"{a}hn, {\it et al.},
  %``Constraints on the high-density nuclear equation of state from the phenomenology of compact stars and heavy-ion collisions,''
  \textit{Phys. Rev.} {\bf C 74} (2006) 035802.

\bibitem{softeos4}
  C.~Fuchs,
  %``Kaon production in heavy ion reactions at intermediate energies,''
  \textit{Prog. Part. Nucl. Phys.}  {\bf 56} (2006) 1.

\bibitem{softeos5}  
 C.~Hartnack, \textit{et al.},
  %``Strangeness Production close to Threshold in Proton-Nucleus and Heavy-Ion Collisions,''
  \textit{Phys. Rept.}  {\bf 510} (2012) 119.
  
\bibitem{babis}
Ch.~Moustakidis, T.~Gaitanos, Ch.~Margaritis, G.A.~Lalazissis,
\textit{Phys. Rev.} \textbf{C 95} (2017) 045801.

\bibitem{puzzle}
I.~Bombaci, arXiV:1601.05339 [nucl-th],\\
D.~Chatterjee, I.~Vidana, 
\textit{Eur. Phys. J.} \textbf{A 52} (2016) 29.

\bibitem{puzzle1}
J.~Haidenbauer, U.~G.~Mei\ss{}ner, N.~Kaiser and W.~Weise,
%``Lambda-nuclear interactions and hyperon puzzle in neutron stars,''
\textit{Eur. Phys. J.} \textbf{A 53} (2017) no.6, 121.
%doi:10.1140/epja/i2017-12316-4
%[arXiv:1612.03758 [nucl-th]].
%31 citations counted in INSPIRE as of 13 Nov 2020

\bibitem{review}
S.~Petschauer, J.~Haidenbauer, N.~Kaiser, U.~G.~Mei\ss{}ner and W.~Weise,
%``Hyperon-nuclear interactions from SU(3) chiral effective field theory,''
\textit{Front. in Phys.} \textbf{8} (2020) 12.
%doi:10.3389/fphy.2020.00012
%[arXiv:2002.00424 [nucl-th]].
%3 citations counted in INSPIRE as of 13 Nov 2020

\bibitem{rmf-others-1}
J.~Schaffner and I.~N.~Mishustin,
%``Hyperon rich matter in neutron stars,''
\textit{Phys. Rev.} \textbf{C 53} (1996) 1416.
%doi:10.1103/PhysRevC.53.1416
%[arXiv:nucl-th/9506011 [nucl-th]].
%399 citations counted in INSPIRE as of 13 Nov 2020

\bibitem{rmf-others-2}
N.~Hornick, L.~Tolos, A.~Zacchi, J.~E.~Christian and J.~Schaffner-Bielich,
%``Relativistic parameterizations of neutron matter and implications for neutron stars,''
\textit{Phys. Rev.} \textbf{C 98} (2018) no.6, 065804.
%doi:10.1103/PhysRevC.98.065804
%[arXiv:1808.06808 [astro-ph.HE]].
%22 citations counted in INSPIRE as of 13 Nov 2020

\bibitem{rmf-others-3}
J.~E.~Christian and J.~Schaffner-Bielich,
%``Twin Stars and the Stiffness of the Nuclear Equation of State: Ruling Out Strong Phase Transitions below $1.7n_0$ with the New NICER Radius Measurements,''
\textit{Astrophys. J. Lett.} \textbf{894} (2020) no.1, L8.
%doi:10.3847/2041-8213/ab8af4
%[arXiv:1912.09809 [astro-ph.HE]].
%22 citations counted in INSPIRE as of 13 Nov 2020

\bibitem{nld}  
T.~Gaitanos, M.~Kaskulov,
  %``Momentum dependent mean-field dynamics of compressed nuclear matter and neutron stars,''
  \textit{Nucl. Phys.} {\bf A 899} (2013) 133,\\
T.~Gaitanos, M.~Kaskulov,
  %``Toward relativistic mean-field description of $\bar{\text{N}}$-nucleus reactions,''
  \textit{Nucl. Phys.} {\bf A 940} (2015) 181.  

\bibitem{rhd}
H.-P.~Duerr, \textit{Phys. Rev.} \textbf{103} (1956) 469,\\
J.D.~Walecka, \textit{Ann. Phys.} \textbf{83} (1974) 491,\\
J.~Boguta, A.~Bodmer, \textit{Nucl. Phys.} \textbf{A 292} (1977) 413.

\bibitem{eft}
S.~Petschauer, \textit{et al.}, 
\textit{Eur. Phys. J.} \textbf{A 52} (2016) 15,\\
J.~Haidenbauer, private communication.

\bibitem{eft-new}
J.~Haidenbauer, U.~G.~Mei\ss{}ner and A.~Nogga,
%``Hyperon\textendash{}nucleon interaction within chiral effective field theory revisited,''
\textit{Eur. Phys. J.} \textbf{A 56} (2020) no.3, 91.
%doi:10.1140/epja/s10050-020-00100-4
%[arXiv:1906.11681 [nucl-th]].
%26 citations counted in INSPIRE as of 13 Nov 2020

\bibitem{LQCD}
T.~Inoue [LATTICE-HALQCD],
%``Hyperon single-particle potentials from QCD on lattice,''
\textit{PoS} \textbf{INPC2016} (2016), 277, \\
%doi:10.22323/1.281.0277
%[arXiv:1612.08399 [hep-lat]].
%10 citations counted in INSPIRE as of 11 Jan 2021
T.~Inoue [HAL QCD],
%``Strange Nuclear Physics from QCD on Lattice,''
\textit{AIP Conf. Proc.} \textbf{2130} (2019) no.1, 020002.
%doi:10.1063/1.5118370
%[arXiv:1809.08932 [hep-lat]].
%8 citations counted in INSPIRE as of 11 Jan 2021
%T.~Inoue (HAL QCD Collaboration), \textit{PoS (INPC2016)} (2016) 277, \\
%T.~Inoue (HAL QCD Collaboration), \textit{AIP Conf. Proc.} (2019) 2130.

\bibitem{nl-omega-1}
M.~Fortin, S.~S.~Avancini, C.~Provid\^encia and I.~Vida\~na,
%``Hypernuclei and massive neutron stars,''
\textit{Phys. Rev.} \textbf{C 95} (2017) no.6, 065803.
%doi:10.1103/PhysRevC.95.065803
%[arXiv:1701.06373 [nucl-th]].
%49 citations counted in INSPIRE as of 23 Nov 2020

\bibitem{nl-omega-2}
C.~Provid\^encia and A.~Rabhi,
%``Hypernuclei and massive neutron stars,''
\textit{Phys. Rev.} \textbf{C 87} (2013) 055801.

\bibitem{nl-omega-3}
Y.~Sugahara, and H.~Toki, \textit{Nucl. Phys.} \textbf{A 579} (1994) 557.

\bibitem{juelich}
J.~Haidenbauer and U.-G.~Meissner, 
\textit{Phys. Rev.} \textbf{C 72} (2005) 044005.
  
\bibitem{Larionov}
A.B. Larionov, \textit{et al.}, 
\textit{Phys. Rev.} \textbf{C 80} (2009) 021601(R).  

\bibitem{nldnuclei}
S.~Antic and S.~Typel,
  %``Relativistic mean-field model with energy dependent self-energies,''
  \textit{AIP Conf. Proc.}  {\bf 1645} (2015) 276.


%%%%%%%%%%%%%%%%%%%%%%%%%%%%%%%%%%%%%%%%%%%%%%%%%%%%%%%%%%%%%%%%%%


\end{thebibliography}

\end{document}